\documentclass[fleqn,twoside]{article}
\usepackage{espcrc2}

\usepackage{graphicx}
\usepackage[figuresright]{rotating}

\newcommand{\ra}{{\rightarrow}}

\newcommand{\gsim}{ \mathop{}_{\textstyle \sim}^{\textstyle >} }

\newcommand{\AmS}{{\protect\the\textfont2
  A\kern-.1667em\lower.5ex\hbox{M}\kern-.125emS}}

\title{Phenomenology of future neutrino experiments with large $\theta_{13}$
}

\author{Hisakazu Minakata\address{Department of Physics, 
Tokyo Metropolitan University, Hachioji, Tokyo 192-0397, Japan}
}

\begin{document}

\begin{abstract}

The question ``how small is the lepton mixing angle $\theta_{13}$?'' had a convincing answer in a surprisingly short time, $\theta_{13} \simeq 9^{\circ}$, a large value comparable to the Chooz limit. It defines a new epoch in the program of determining the lepton mixing parameters, opening the door to search for lepton CP violation of the Kobayashi-Maskawa-type.
I discuss influences of the large value of $\theta_{13}$ to search for CP violation and determination of the neutrino mass hierarchy, the remaining unknowns in the standard three-flavor mixing scheme of neutrinos. I emphasize the following two points: (1) Large $\theta_{13}$ makes determination of the mass hierarchy easier. It stimulates to invent new ideas and necessitates quantitative reexamination of practical ways to explore it. (2) However, large $\theta_{13}$ does not quite make CP measurement easier so that we do need a ``guaranteeing machine'' to measure CP phase $\delta$.

\end{abstract}

\maketitle

\section{A year from June 2011: The year of $\theta_{13}$}

This is the memorable conference in this series in which everybody was convinced that the mixing angle $\theta_{13}$, the only angle that remained unknown in the three-flavor lepton mixing \cite{MNS},\footnote{
For history of discussions of lepton flavor mixing in the early days, see \cite{Kobayashi_nu2012}.  
}
%
is nonzero and is determined already in an unexpected high precision. 
Remarkably, the drama, which has been started as T2K announcement of six electron appearance events \cite{T2K}, had a beautiful finale within a year! In addition to a followup by MINOS \cite{MINOS}, it is largely due to the contribution by the three dedicated reactor $\theta_{13}$ experiments \cite{DC,DB,RENO}.

Then, what was new in this conference? To my understanding they are:

\begin{itemize}

\item 
By the updated data sets from T2K \cite{T2K_nu2012} and Double Chooz \cite{DC_nu2012}, the confidence level (CL) for nonzero $\theta_{13}$ becomes greater than 3$\sigma$ in each individual experiment.

\item 
With their updated data Daya Bay alone achieved 7.7$\sigma$ CL for nonzero $\theta_{13}$ \cite{DB_nu2012}, the same CL as reached by the combined analysis of all the five (reactor and accelerator) experiments before this conference \cite{MMNZ11}.

\end{itemize}

\noindent
Clearly the extremely high accuracy of measurement achieved by the reactor experiments is remarkable. It is partly due to controlled small systematic errors realized by the near-far identical multi-detector method \cite{Mikaelyan,MSYIS}. 
I am sure that the new and powerful contributions from East Asia \cite{DB,RENO,DB_nu2012} to neutrino experiments are highly welcomed by people in this community 

I want to emphasize, however, that accelerator and reactor measurement result in a consistent value of $\theta_{13}$, within errors, is equally important, because the framework of three neutrino mixing itself has never been tested in a high precision. To meet this goal reducing the (in particular statistical) errors in accelerator measurement is highly needed. 

It may be worth to point out that we hitherto enjoyed redundancy in determination of the lepton mixing parameters: $\Delta m^2_{32}$ and $\theta_{23}$ are determined independently by the atmospheric \cite{SK_nu2012} and the accelerator \cite{MINOS_nu2012} experiments.\footnote{
Notice that I am not talking about priority issue here. That is the reason why I quote only the latest references.
}
%
(The former discovered neutrino oscillation \cite{Kajita-discovery}, the phenomenon discussed in \cite{Pontecorvo}.) 
Similarly, $\Delta m^2_{21}$ and $\theta_{12}$ are measured by the solar \cite{solar_nu2012} and the KamLAND \cite{Gando} experiments independently.
In this sense, it is great that we were able to maintain double coverage of $\theta_{13}$ determination by the reactor and the accelerator experiments.

With all the data set including the solar, reactor, atmospheric, and the accelerator data $\theta_{13}$ are constrained as \cite{concha} (see also \cite{bari})
\begin{eqnarray}
\sin^2 2 \theta_{13}= 0.090^{+0.008 (0.027)}_{-0.010 (0.029)}
\label{theta13-global}
\end{eqnarray}
where the errors outside and in parentheses are with $1\sigma$ and $3\sigma$ CL, respectively.

The fact that $\theta_{13}$ is relatively large is entirely natural as I argued in my Nu2008 talk in Christchurch. I quote here what I wrote \cite{Nu2008-mina}: 
``An interesting question is how large is $\theta_{13}$. In this talk, I argue that it must be large. Since the MNS matrix is the product of two matrices which diagonalize the lepton and the neutrino mass matrices, one out of three angles cannot be too small given the fact that the remaining two are large.'' 

\section{What is next?}

We now know values of all the mixing angles of the MNS matrix, up to experimental uncertainties: $\theta_{12} \simeq 34^{\circ}$, $\theta_{23} \sim 45^{\circ}$, $\theta_{13} \simeq 9^{\circ}$. Therefore, the answer to the question ``What is next?'' is obvious; 
Lepton CP violation and the neutrino mass hierarchy, the only remaining unknowns in the standard three-flavor lepton mixing\footnote{
I regret that I was not able to cover important topics such as absolute neutrino masses, zero-$\nu$ double beta decay, nature of neutrinos (Majorana vs. Dirac) and if Majorana, the issue of associated CP violating phases.
}

Why so much interests in the lepton CP phase $\delta$ and the mass pattern of neutrinos? To my prejudice leptons and quarks must be related with each other at some deeper level. In the Standard Model (SM) the relationship manifests itself through quantum anomaly: the theory fails without quarks, or without leptons. With anomaly being a short distance effects I believe that it allows interpretation as evidence for the relationship between leptons and quarks at a level much deeper than the SM.

If this view has anything to do with nature it is important to know every aspect of the question of to what extent they look similar or different from each other. Therefore, we would like to know whether the Kobayashi-Maskawa mechanism \cite{KM} for CP violation in the quark sector prevails to the lepton sector, or if it is the unique source for CP violation. Since our theoretical expectations like ``flavor mixing angle has to be small'' badly failed for leptons, we have to be free from any prejudices such as ``the neutrino mass pattern must be resemble that of charged leptons''. We need to know the answer to these questions experimentally. 

\section{``All in one'' approach}

What is the greatest impact of large $\theta_{13}$ on future strategies for exploring the remaining unknowns? To my opinion they are the following three points:

\begin{itemize}

\item 
It appears that people's opinion converged about what would be the next step, namely, conventional superbeam \cite{superbeam}.\footnote{
I would like to note, however, that other options, e.g., neutrino factory \cite{nufact} or beta beam \cite{beta} would have chance of playing crucial roles in future experiments. For example, if muon collider is on our way neutrino factory would be a natural step toward such a novel beam technology. 
}

\item 
The most important point to me is that large $\theta_{13}$ allows us to take ``all in one'' approach \cite{Nufact11-mina}: 
%
A single apparatus such as Hyper-Kamiokande (Hyper-K) \cite{HK-LOI} (even though its baseline is relatively short) or LBNE \cite{LBNE} can become a ``guaranteeing machine'' for CP and the mass hierarchy.

\item 
With large $\theta_{13}$ the mass hierarchy could be determined by various ways even before a ``guaranteeing machine'' is constructed, allowing multiple coverage of its determination.

\end{itemize}

\noindent
Let me start by explaining in the next section why only the mass hierarchy, not CP, becomes easier with large $\theta_{13}$. 

\section{How and to what extent large $\theta_{13}$ changes things}
\label{how-change}

Here, I address the question of whether or to what extent a large value of $\theta_{13} \simeq 9^\circ$ really helps discovery of CP violation and determination of the mass hierarchy. I argue that the answer is: ``{\em definitely yes} for the mass hierarchy, but not quite for CP''.\footnote{
Of course, it is true that if $\theta_{13}$ is extremely small measuring CP violating phase would have been formidable. In this sense the large $\theta_{13}$ opened the possibility of accessing to the fascinating phenomena, as emphasized by Nishikawa-san \cite{Nishikawa}. Sharing (of course!) the same bottom-line understanding, what I would like to discuss here is a more practical questions that people may face in reality. }

\subsection{CP vs. mass hierarchy}

To show the point I appeal to a well known expression of $\nu_\mu \ra \nu_e$
appearance probability, which is known as the Cervera {\it et al.} formula \cite{golden}:  
\begin{eqnarray}
&&P(\nu_{\mu} \ra \nu_e) 
\nonumber \\
&=& 
4 \frac{ ( \Delta m^2_{31} )^2 }{  (  \Delta m^2_{31} - a )^2  } 
s^2_{23} s^2_{13} \sin^2 \left(\frac{ ( \Delta m^2_{31} - a) L}{4E}  \right)  
\nonumber \\
&+& 8 J_r \frac { \Delta m^2_{31}  \Delta m^2_{21}  }{  a ( \Delta m^2_{31} - a )} 
\sin \left(\frac{  a L}{4E}  \right) 
\nonumber \\
&\times&
\sin \left(\frac{ ( \Delta m^2_{31} - a ) L}{4E}  \right)  
\cos \left(  \delta +  \frac{  \Delta m^2_{31} L}{4E}   \right) 
\nonumber \\
&+&  
4 \left( \frac{ \Delta m^2_{21} } { a } \right)^2 
c^2_{12} s^2_{12} c^2_{23} 
\sin^2 \left(\frac{ a L}{4E}  \right).  
\label{Pmue-2nd}
\end{eqnarray}
where $\Delta m^2_{ij} \equiv m^2_i - m^2_j$, $J_r \equiv c_{12} s_{12} c_{23} s_{23} s_{13}$, and $a \equiv 2\sqrt{2} G_F N_e(x) E$ is a coefficient related to neutrino's refraction in medium of electron number density $N_e(x)$~\cite{wolfenstein}, where $G_F$ is the Fermi constant and $E$ is the neutrino energy. 

The first term in (\ref{Pmue-2nd}) can be called the ``atmospheric oscillation'' term, and the third one ``solar oscillation'' term. The second term is an interference term, the unique one which carries information of CP phase $\delta$.
The atmospheric oscillation term is proportional to $s_{13}^2$, and this is the term most sensitive to the matter effect. Therefore, resolution of the mass hierarchy has a great advantage of the large $\theta_{13}$.

Numbers of electron appearance event increases as $\theta_{13}$ becomes large, so that naively it would merit CP measurement. However, the large atmospheric term acts as a ``background'' in measurement of $\delta$. In fact, the CP asymmetry is inversely proportional to $\theta_{13}$: 
$A_{CP} \equiv \frac{P(\nu_{\mu} \ra \nu_e) - P(\bar{\nu}_{\mu} \ra \bar{\nu}_e)}{P(\nu_{\mu} \ra \nu_e) + P(\bar{\nu}_{\mu} \ra \bar{\nu}_e)} \propto 1/\theta_{13}$. Competition of these contradictory factors makes CP sensitivity roughly independent of $\theta_{13}$ in region $\sin^2 2\theta_{13} \gsim 0.4$, as seen e.g., in LOI of Hyper-K experiment \cite{HK-LOI}.

\subsection{Consistent argument?}

The readers may wonder if my above discussion is consistent. The expression (\ref{Pmue}) of $P(\nu_{\mu} \ra \nu_e)$ can be derived as a second-order ($\sim\epsilon^2$) perturbative formula with the small expansion parameter $\epsilon \equiv \frac{\Delta m^2_{21}}{\Delta m^2_{31}} \simeq 0.03$, under the assumption $s_{13} \sim \epsilon$. However, the measured value of $\theta_{13}$ is large, $s_{13} \simeq \sqrt{ \epsilon } = 0.18$ rather than $s_{13} \sim \epsilon$. Therefore, there must exist higher order terms in $s_{13}$ which could invalidate my argument in the previous subsection. 

The problem of higher order correction to order $s_{13}^4 \sim \epsilon$ to the Cervera {\it et al.} formula has been investigated in \cite{Large13-P}. The answer to the above question is: Fortunately, my argument is still valid qualitatively. 
The point is that the higher order correction terms in $s_{13}$ do not have $\delta$ dependence to order $\epsilon^2$. This property follows from a general theorem which states that: 

\begin{itemize}

\item 
$\delta$-dependent terms in the oscillation probability in matter, 
not only $\sin \delta$ but also $\cos \delta$ terms, must 
come with the two suppression factors, 
$\frac{ \Delta m^2_{21} }{ \Delta m^2_{31} }$ and 
the reduced (or ``not quite'') Jarlskog coefficient 
$J_{r} \equiv c_{12} s_{12} c_{23} s_{23} s_{13}$.

\item 
In $\nu_e$-related oscillation probabilities the $\delta$-dependence 
exists only in odd terms in $s_{13}$. Conversely, all the odd terms in 
$s_{13}$ are accompanied with either $\cos \delta$ or $\sin \delta$.

\end{itemize}

\noindent
Because of the theorem $\delta$-dependent higher-order correction starts at order $\epsilon^{5/2} \sim \mathcal{O} (10^{-4})$. For more details, see \cite{Large13-P}.

In what follows I would like to discuss possible ways for determination of the mass hierarchy and CP phase $\delta$ one by one. 

\section{Mass hierarchy determination: ``hundred flowers campaign''}

Thanks to the large value of measured $\theta_{13}$ ``hundred ideas'' started to show up, or obtained renewed interests.
%
Below I will try to describe some of them. Notice that the order of description I choose reflects neither feasibility nor robustness of the proposed measurement. I have to apologize if any relevant ideas are missed in my following discussion. Even worse, you may find that my assessment is not equal to everyone, too critical to A and too warm for B. But, this is my honest feeling (or understanding) at the present stage.  

I expect that with large $\theta_{13}$ the question ``Who determines mass hierarchy first, say at $3\sigma$ CL?'' will remain the most tantalizing one in coming $\sim10$ years.

\subsection{Lucky NO$\nu$A and/or lucky T2K}

\begin{figure}[htb]
\begin{center}
\includegraphics[width=16pc]{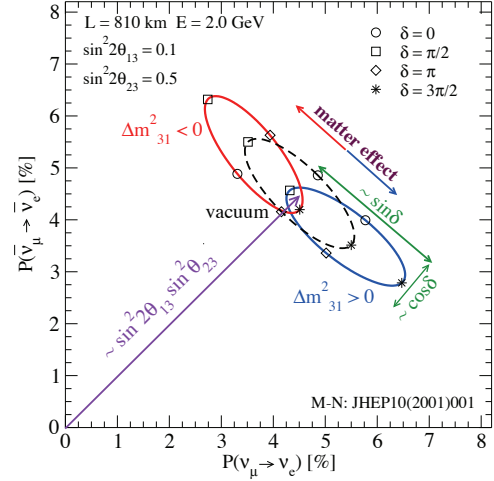}
\vspace{-4mm}
\caption{
A $P(\nu_\mu \rightarrow \nu_e) - P(\bar{\nu}_\mu \rightarrow \bar{\nu}_e)$ 
bi-probability plot with experimental parameters corresponding to NO$\nu$A experiment is presented to exhibit characteristic features of the neutrino oscillations. It can display, in a compact fashion, the sizes of competing three effects, CP violating and CP conserving effects due to $\delta$ as well as the matter effects \cite{MNJhep01}.
}
\label{bi-P}
\end{center}
\vspace{-6mm}
\end{figure}

It would be nice if the mass hierarchy can be resolved by the existing facilities. It can happen with some (extreme or modest?) luckiness. To show the point let me draw the bi-probability plot in $P - \bar{P}$ [$P \equiv P(\nu_\mu \rightarrow \nu_e), \bar{P} \equiv P(\bar{\nu}_\mu \rightarrow \bar{\nu}_e)$] space in Fig.~\ref{bi-P} \cite{MNJhep01}. If $\delta \simeq 3\pi/2$ ($\pi/2$), the asymmetry between $P$ and $\bar{P}$ 
is largest for the normal (inverted) hierarchy. Therefore, the mass hierarchy can be resolved if the statistics is sufficient. This is what happens in the often showed NO$\nu$A sensitivity plot \cite{NOVA}. With extreme statistics the same thing can occur in T2K.

\subsection{NO$\nu$A+INO} 

If we cannot enjoy the luckiness we must think of some help from other experiments. India-based Neutrino Observatory (INO) \cite{INO}, which is funded, is a natural candidate among the possible list. INO can observe atmospheric neutrinos (see the next section for more about it) and has capability of distinguishing $\nu$ and $\bar{\nu}$ by magnetized detector. This possibility was examined in detail in \cite{Blennow-Schwetz}. The result indicates that the mass hierarchy resolution may be possible at $3 \sigma$ CL for all values of $\delta$ if the systematic errors are under control to $\sim10\%$ level. 

\subsection{CERN$\rightarrow$Super-K} 

\begin{figure}[htb]
\begin{center}
\includegraphics[width=16pc]{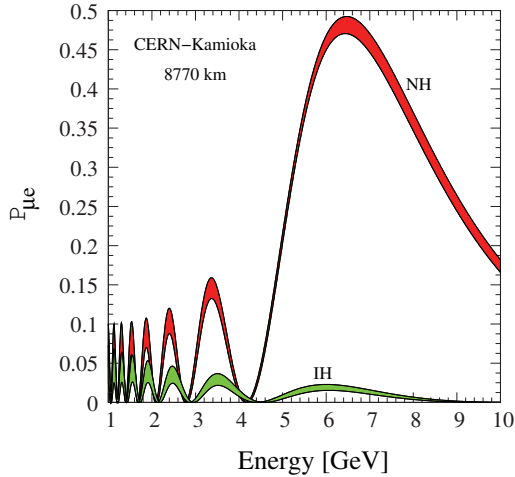}
\caption{
Presented is $P_{\mu e} \equiv P(\nu_\mu \rightarrow \nu_e)$ as a function of $E$, which is taken from \cite{Agarwalla-Hernandez}. The large asymmetry between $P_{\mu e}$ of the two hierarchies reflects the feature that the blue and the red ellipses in Fig.~\ref{bi-P} become far apart at the super-long baseline. The band width, which represents the effect of $\delta$, is so small because the ellipses shrink at high energies. 
}
\label{Pmue}
\end{center}
\vspace{-4mm}
\end{figure}

The authors of \cite{Agarwalla-Hernandez} propose to construct a CERN beam line which points toward the existing detector Super-Kamiokande with 22.5kt of fiducial mass in Japan. With the baseline of 8770 km and a 50 GeV proton beam as intense as $3 \times 10^{21}$ POT the mass hierarchy resolution at $5 \sigma$ CL is possible just by counting the number of events in the neutrino mode. This exercise demonstrates that with sufficiently long baseline (and with reasonable number of events) mass hierarchy resolution is a simple matter. It may be understood by looking into Fig.~\ref{Pmue}.

\section{Mass hierarchy resolution by atmospheric neutrinos} 

The above consideration strongly suggests that by utilizing atmospheric neutrinos (some portion of which cross the core of the Earth) the mass hierarchy resolution can be done without using accelerator beam. Certainly, this is not a new idea and has been discussed by many authors in the past. (Sorry, too many works to quote here.) See e.g., \cite{peres-smirnov03,bernabeu03} for early references. With large $\theta_{13}$ it is the time to revisit the idea and carry out quantitative evaluation of its sensitivity to the mass hierarchy determination. A warning is that it has been known that a huge number of events is required to make it happen \cite{Kajita-NOON04}. 

I mention here a few particular examples. 

\subsection{PINGU} 

PINGU (Phased Ice-cube Next Generation Upgrade) \cite{PINGU} is a proposal of constructing even denser core (with $\simeq 20$ strings) in the Deep Core in the IceCube detector to lower the threshold to a few GeV. Because of huge detector volume of several megaton it can collect gigantic numbers of atmospheric neutrino events. 

\begin{figure}[htb]
\begin{center}
\includegraphics[width=18pc]{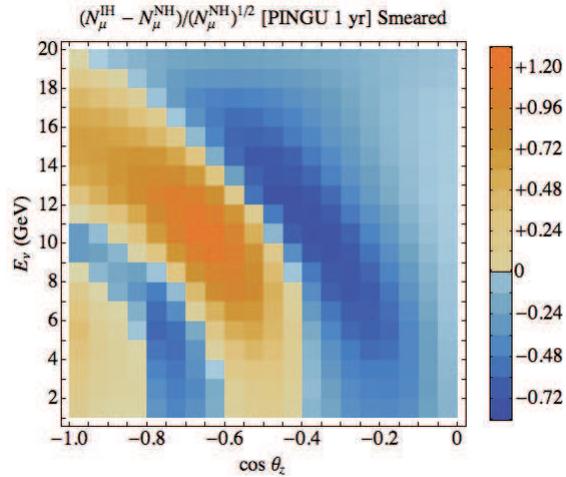}
\vspace{-8mm}
\caption{
The mass hierarchy asymmetry calculated with $\nu_\mu$ events in the reconstructed energy and zenith angle ($E^r_\nu$--$\cos\theta^r$) plane.  Gaussian smearing over the bins is performed with widths 
$\sigma_E = 0.2E_\nu$ and $\sigma_\theta = \sqrt{m_p/E_\nu}$. For details see \cite{Akhmedov2012} from which this figure is taken.
}
\label{MH-asymmetry}
\end{center}
\vspace{-2mm}
\end{figure}

The question of which region of energy and zenith angle is most sensitive to the mass hierarchy (and $\delta$ and $\theta_{23}$) is investigated in detail in \cite{Akhmedov2012}. See Fig.~\ref{MH-asymmetry}. After taken into account the energy and angle resolutions they concluded that the mass hierarchy resolution can be possible at $(4-11)\sigma$ CL with 5 years operation of PINGU. In view of the relatively low cost (20-30 million dollars as quoted) of the detector upgrade in IceCube, it may be one of the leading competitors in the ``mass hierarchy race''.

\subsection{Megaton water Cherenkov or 100 kiloton liquid Ar} 

Of course, PINGU is not the unique candidate for those who can enjoy huge number of atmospheric neutrino events. Any large volume apparatus qualifies for this purpose. In fact, a detailed analysis of mass hierarchy sensitivity by water Cherenkov detector is carried out by the Hyper-K collaboration with the fiducial mass of 0.56 megaton \cite{HK-LOI}. The result indicates that 
(1) the sensitivity to the mass hierarchy depend very much on the value of $s_{23}^2$ (better for larger $s_{23}^2$), 
(2) it can be carried out by $\sim 3-10$ years running of Hyper-K for any allowed values of $s_{23}^2$. 

Similarly, $\sim100$ kiloton scale liquid Ar detector can achieve a comparable sensitivity \cite{Barger-etal}. 

\section{Mass hierarchy resolution in vacuum}

The mechanisms discussed so far by which the mass hierarchy can be distinguished utilize the neutrino matter interactions. I note that this is {\em not} the only way to determine the mass hierarchy; It can be done in vacuum!; There are (at least) two options and large $\theta_{13}$ helps in both cases. 

\subsection{Sign[$\Delta m^2_{atm} (ee) - \Delta m^2_{atm} (\mu\mu)$] method} 

This is a simpler one, at least at the conceptual level. One can easily show that 
the difference $\Delta m^2_{atm} (ee) - \Delta m^2_{atm} (\mu\mu)$ between the two $\Delta m^2_{atm}$ measured in $\nu_e$ and $\nu_\mu$ disappearance channels is positive (negative) if the mass hierarchy is normal (inverted) \cite{NPZ,MNPZ1}. A serious problem in this method is that an accurate measurement of $\Delta m^2_{atm}(ee)$ and $\Delta m^2_{atm} (\mu\mu)$ at sub-\% level is required. Unfortunately, no practical way is known to achieve such accuracy, as discussed in \cite{MNPZ1}. 

\subsection{$\Delta m^2_{32}$ and $\Delta m^2_{31}$ oscillations phase difference method} 

If one can measure neutrino energy spectrum in an excellent energy resolution to a few \% at the baseline where solar-scale oscillation is large, one would find wiggles due to atmospheric scale oscillation superposed onto the long wavelength solar oscillations. Depending upon the mass hierarchy, the phase of the atmospheric-scale oscillation is advanced (normal) or retarded (inverted). This possibility was first raised in \cite{petcov}. The key for this method to work, necessity of observing wiggles over many periods with high accuracy, was recognized in \cite{hanohano,DB2} who used the Fourier transform method. 

A way to understand the phenomenon in a clear cut manner is to write $P(\nu_e \rightarrow \nu_e)$ in the following form \cite{MNPZ2} with $P_{\odot}$ being the solar oscillation term:  
\begin{eqnarray}
&& P(\nu_e \rightarrow \nu_e) + P_{\odot} = 1-
\frac{1}{2} \sin^2 2\theta_{13} \times 
\nonumber \\
&&\left[ 1- \sqrt{1- \sin^2 2\theta_{12} \sin^2 \Delta_{21} }
~\cos (2 \Delta_{ee} \pm \phi_\odot) \right] \nonumber
\label{Pee}
\end{eqnarray}
where 
$\Delta_{ji} \equiv \vert \Delta m^2_{ji} \vert L/4E$, 
$\Delta m^2_{ee} \equiv \alpha \vert \Delta m^2_{32} \vert + (1-\alpha) \vert \Delta m^2_{31} \vert$ ($0 \leq \alpha \leq 1$).\footnote{
In \cite{MNPZ2} $\alpha=s^2_{12}$ is used in which case $\Delta m^2_{ee}$ is an effective $\Delta m^2_{atm}$ parameter which is measured in $\nu_e$ disappearance channel \cite{NPZ}. But, the particular form is not essential and the case with generic $\alpha$ does the job.
}
%
$\phi_\odot$ defined by 
\begin{eqnarray}
\sin \phi_\odot = 
\frac{ c^2_{12} ~\sin ( 2 \alpha \Delta_{21})
- s^2_{12}~ \sin ( 2 (1-\alpha) \Delta_{21} ) }
{\sqrt{1-\sin^2 2 \theta_{12} \sin^2 \Delta_{21}}} \nonumber 
\label{phi-def}
\end{eqnarray}
represents effect of phase advancement or retardation; 
The positive (negative) {\it sign} in front of $\phi_\odot$ corresponds to the normal (inverted) mass hierarchy, and the phase difference shows up at baseline $\phi_\odot \simeq O(1)$. 
In the kinematic regions where $\phi_\odot$ is constant the mass hierarchy can be confused by taking $2 \Delta_{ee} (\mbox{NH}) + \phi_\odot = 2 \Delta_{ee} (\mbox{IH}) - \phi_\odot$ which can be done well within the expected uncertainties of $\Delta m^2_{ee}$ \cite{MNPZ2,Parke-NOW08}. 
 
This difficulty can be circumvented by collecting informations of phases over many wiggles. An excellent relative energy resolution is then required. Recently, by integrating all these discussions, a group of reactor experimentalists made a careful examination which produced a complete list of experimental requirements for this method to work \cite{Qian-etal}. Certainly, separating two $\Delta m^2_{atm}$ oscillations is a fascinating but a challenging goal.

\section{Lepton CP phase $\delta$}

As I argued in section~\ref{how-change}, large $\theta_{13}$ does not quite help measurement of the CP phase $\delta$, as can be confirmed by looking into Figures~21-23 of Hyper-K sensitivity study \cite{HK-LOI}. An operation of J-PARC-HK of $\simeq1$MW$\cdot$year is required to obtain a reasonable sensitivity. It is roughly equivalent to $\sim50$ years of SK operation assuming J-PARC beam intensity of 400 kW, which is neither practical nor wanted. Therefore, unlike the case of mass hierarchy determination, we do need a dedicated apparatus that can assure the detection of CP phase $\delta$. This point, the absolute necessity for ``guaranteeing machine'' for CP, was the point of the strongest emphasis in my talk. 

One may ask: Should we wait until completion of such dedicated machines and their measurement which would take at least 10 years from now? I think it a very relevant questions considering the length of time needed. Below, I describe some possibilities. 

\section{Accelerator-reactor method and use of atmospheric neutrinos for CP}

It was proposed \cite{reactor-CP} that combining $\delta$-sensitive superbeam experiments with $\delta$-independent measurement by reactor $\theta_{13}$ experiments one can obtain the information of CP phase, which however practically limited to the sign$(\sin \delta)$. An analysis with the current data is given in \cite{MMNZ11}, from which, of course, one cannot say much. 

\begin{figure}[htb]
\begin{center}
\includegraphics[width=20pc]{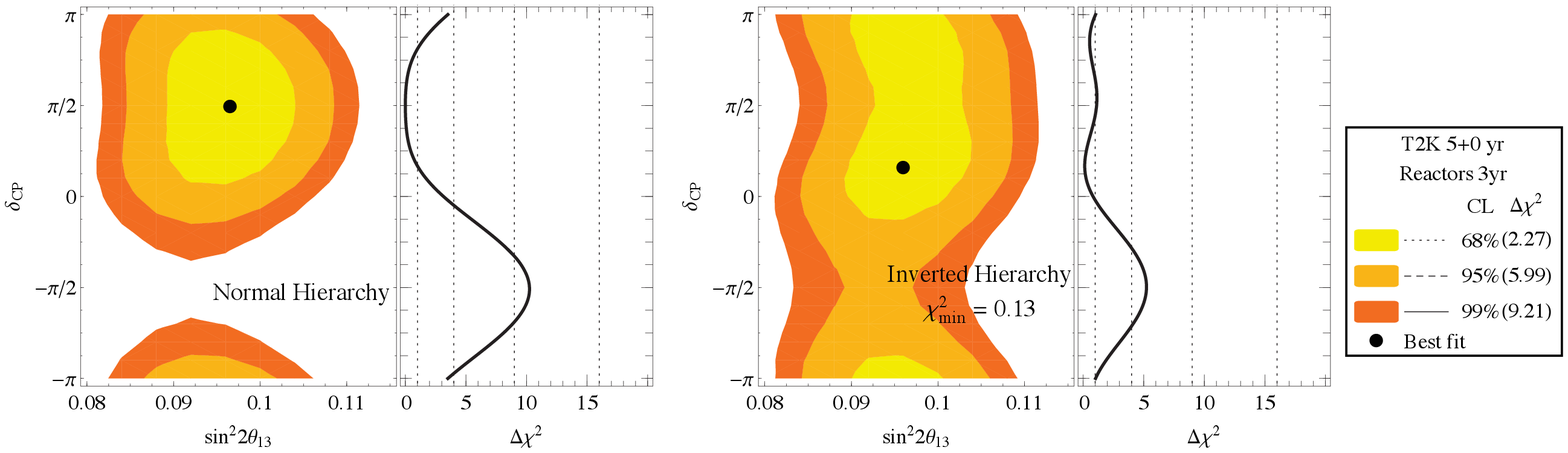}
\includegraphics[width=20pc]{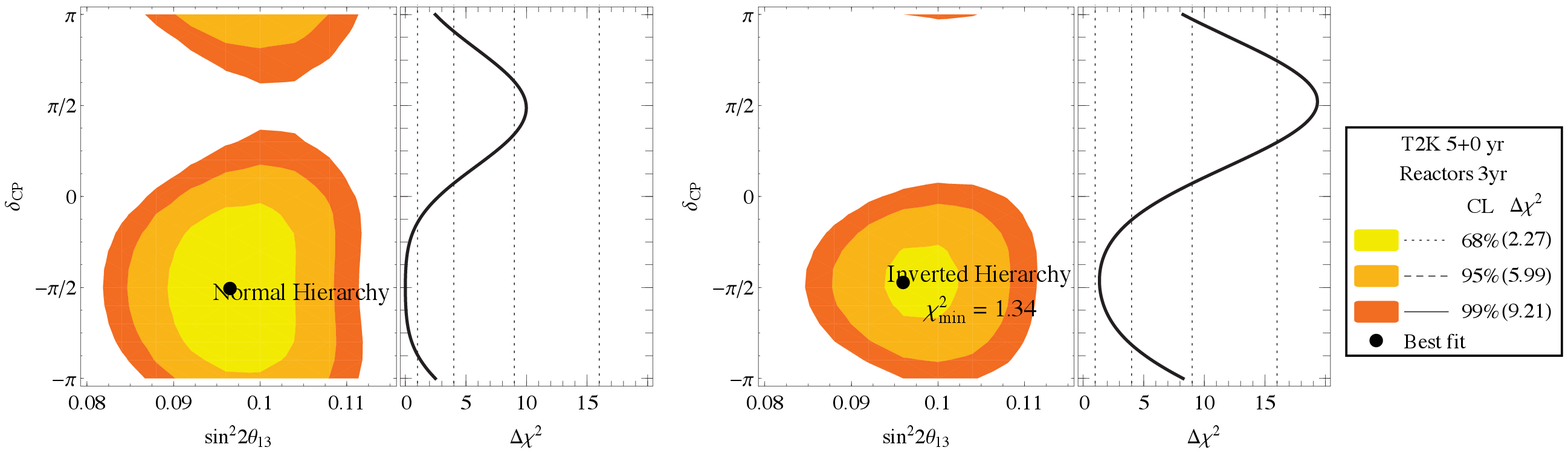}
\vspace{-8mm}
\caption{
The region in $\sin^2 2\theta_{13} - \delta$ space allowed by 5 years measurement of T2K, corresponding to $5 \times 10^{21}$ POT in its neutrino mode, and 
3 years operation of all the three reactor $\theta_{13}$ experiments. 
For T2K 10\% systematic errors and 1\% determination of $\sin^2 2\theta_{23}$ are assumed. }
\label{accelerator-reactor}
\end{center}
\vspace{-8mm}
\end{figure}

In this context 
one of the interesting questions to ask is: 
``To what extent CP phase $\delta$ can be constrained by, e.g., T2K neutrino run of 5 years and the high precision data from reactor experiments for 3 years?''. These questions are under investigation with updated fluxes and cross sections of the T2K experiment \cite{MMNZ12}. Figure~\ref{accelerator-reactor}, which is an updaded version of the one presented in poster presentation in Nu2012, gives a partial answer to this question for input $\delta = \pi/2$, or $\delta = 3\pi/2$ taking the normal hierarchy as input. As it stands we cannot claim that $\delta$ can be measured by the accelerator-reactor method. But, it is encouraging to see that sign$(\sin \delta)$ could be determined to a limited CL of $\sim 1\sigma$.
 
Another approach is to use atmospheric neutrinos to obtain informations on $\delta$ 
\cite{takeuchi-Nu2010,SK_nu2012,concha,bari}. However, it appears to be too premature to draw conclusion on which region of $\delta$ is preferred. 

\section{Conclusion}

The record-fast development of $\theta_{13}$ experiments entailed a large measured value of $\theta_{13}$ which is comparable to the Chooz limit. It defines a new stage of our understanding of lepton flavor mixing and opened the exciting possibility of accessing to lepton CP violation and the neutrino mass hierarchy. I tried to discuss various ideas and some relevant issues of future programs for determination of the lepton mixing parameters. 

I concluded my talk by speculating about a fictitious slide to be displayed in Neutrino 202$N$ conference, which says that ``We now know all the $\nu$SM parameters, $\theta_{12} = 34^{\circ}$, $\theta_{23} = 43^{\circ}$, $\theta_{13} = 9^{\circ}$, $\delta \simeq 270^{\circ}$, the mass hierarchy = inverted''. I hope that we will be able to know $N$ soon.

\section*{Acknowledgments}
I thank Stephen Parke for illuminating discussions on the phase difference method for the mass hierarchy resolution.
This work was supported in part by KAKENHI, Grant-in-Aid for Scientific
Research No. 23540315, Japan Society for the Promotion of Science.

\end{document}